\documentclass{ws-procs961x669}            
\usepackage{color}

\begin{document}
\title{A look inside Feynman's route to gravitation}

\author{M. Di Mauro}

\address{Dipartimento di Matematica, Universit\`a di Salerno,
Via Giovanni Paolo II,\\
Fisciano (SA), 84084, Italy\\
E-mail: madimauro@unisa.it}

\author{S. Esposito$^*$ and A. Naddeo$^{**}$}

\address{INFN, Sezione di Napoli, C. U. Monte S. Angelo, Via Cinthia,\\
Napoli, 80125, Italy\\
$^*$E-mail: sesposit@na.infn.it\\
$^{**}$E-mail: anaddeo@na.infn.it}

\begin{abstract}
In this contribution we report about Feynman's approach to
gravitation, starting from the records of his interventions at the
Chapel Hill Conference of 1957. As well known, Feynman was
concerned about the relation of gravitation with the rest of
physics. Probably for this reason, he promoted an unusual, field
theoretical approach to general relativity, in which, after the
recognition that the interaction must be mediated by the quanta of
a massless spin-$2$ field, Einstein's field equations should
follow from the general properties of Lorentz-invariant quantum
field theory, plus self-consistency requirements. Quantum
corrections would then be included by considering loop diagrams.
These ideas were further developed by Feynman in his famous
lectures on gravitation, delivered at Caltech in 1962-63, and in a
handful of published papers, where he also introduced some field
theoretical tools which were soon recognized to be of general
interest, such as ghosts and the tree theorem. Some original
pieces of Feynman's work on gravity are also present in a set of
unpublished lectures delivered at Hughes Aircraft Company in
1966-67 and devoted primarily to astrophysics and cosmology. Some
comments on the relation between Feynman's approach to gravity and
his ideas on the quantum foundations of the fundamental
interactions are included.
\end{abstract}


\bodymatter

\section{Introduction: a timeline}\label{sec1}
Among the many scientific interests that Feynman had in the 1950s
and in the 1960s, a prominent place was taken by the understanding
of the relation of gravitation to the rest of
physics,\footnote{``Next we shall discuss the possible relation of
gravitation to other forces. There is no explanation of
gravitation in terms of other forces at the present time. It is
not an aspect of electricity or anything like that, so we have no
explanation. However, gravitation and other forces are very
similar, and it is interesting to note analogies.'' (Ref.
\citenum{Feynman:1963uxa}, Vol. 1, Sec. 7-7); ``My subject is the
quantum theory of gravitation. My interest in it is primarily in
the relation of one part of nature to another.'' (Ref.
\citenum{Feynman:1963ax}, p. 697)} and in particular the
assessment of its consistency with the uncertainty
principle.\footnote{``...it would be important to see whether
Newton's law modified to Einstein's law can be further modified to
be consistent with the uncertainty principle. This last
modification has not yet been completed.''(Ref.
\citenum{Feynman:1963uxa}, Vol. 1, Sec. 7-8)} Feynman likely began
to seriously think about gravity in the early 1950s, just after
completing his work on quantum electrodynamics. This was attested
by Murray Gell-Mann \cite{Gell-Mann} in a paper written for a
Physics Today special issue devoted to Feynman's
legacy,\footnote{Indeed here Gell-Mann remembered his visit to
Caltech during the Christmas vacation of 1954-55, the discussions
with Feynman about quantum gravity, and the fact that Feynman
``had made considerable progress''.} as well as by Bryce S. DeWitt
\cite{ChapelHill}, in a letter to Agnew Bahnson written in
November 1955 \cite{Letter}. Feynman's efforts began in a period
in which general relativity, after a stagnation which lasted about
thirty years, was gradually emerging as a mainstream research
area, giving rise to a process known as the \emph{renaissance} of
general relativity \cite{BlumLalliRenn}.

A crucial event to consider, in order to reconstruct Feynman's
approach to gravitation, is the 1957 Chapel Hill conference
\cite{ChapelHill} on \emph{The Role of Gravitation in Physics},
which was also pivotal to the renaissance of general relativity.
At that conference, the gravitational physics community delineated
the tracks along which subsequent work would develop in the
subsequent decades. The main threads were the following
\cite{BergmannRMP}: classical gravity, quantum gravity, and the
classical and quantum theory of measurement (as a bridge between
the previous two topics). In that conference Feynman's work on
gravity, of which nothing had been published yet, was presented
for the first time, and put on paper in the records
\cite{ChapelHill}. In fact these written records testify that by
the time of the conference he had already deeply thought, and
performed computations, about each of the above listed three
topics, focusing in particular on classical gravitational waves,
on the arguments in favor of quantum gravity from fundamental
quantum mechanics, and finally on quantum gravity itself. Thus, in
this contribution, we take as starting point for our
reconstruction Feynman's interventions at Chapel Hill and follow
the development of his work in the subsequent years, until the
late sixties, where he apparently lost interest in the subject.

The ultimate goal of Feynman's work was the development of a
quantum field theory of gravitation, which led him to face deep
conceptual as well as mathematical issues, such as divergent
integrals and the lack of unitarity beyond the tree level
approximation. This task accompanied Feynman for some years, as
stated in a letter \cite{WeisskopfLetter} he wrote to Viktor F.
Weisskopf in 1961 (``As you know, I am studying the problem of
quantization of Einstein's General Relativity. I am still working
out the details of handling divergent integrals which arise in
problems in which some virtual momentum must be integrated
over''), as well as reported by William R. Frazer \cite{LaJolla}
in a short summary of the talks given at the La Jolla conference,
later in the same year.\footnote{The International Conference on
the Theory of Weak and Strong Interactions was held in June 14-16,
1961, at the University of California, San Diego, in La Jolla.
Here, Geoffrey F. Chew gave his celebrated talk on the $S$-matrix
\cite{Chew}, while an afternoon session was devoted to the theory
of gravitation, where Feynman reported on his work on the
renormalization of the gravitational field, and recognized
non-unitarity as the main difficulty, which was shared also by
Yang-Mills theory.} A first comprehensive account of Feynman's
results on quantum gravity can be found in the talk he gave in
1962 at the Warsaw conference,\footnote{The International
Conference on General Relativity and Gravitation was held in Jab\l
onna and Warsaw in July 25-31, 1962, with Leopold Infeld as the
chairman of the local organizing committee. The discussion focused
on three main topics: the general properties of gravitational
radiation, the quantization of gravity and the exact solutions of
the Einstein field equations.} whose written version was later
published as a regular paper in Acta Physica Polonica
\cite{Feynman:1963ax}. Further details were given by Feynman much
later, in a couple of papers \cite{WheelerFest1,WheelerFest2}
published in 1972 in the Festschrift for John A. Wheeler's 60th
birthday \cite{Klauder:1972je}, which were written in a period in
which he was already deeply absorbed in the study of partons and
strong interactions.

Among the main sources which contribute to offer a clear account
of Feynman's work on gravity issues, it is mandatory to include
the famous Caltech Lectures on Gravitation \cite{Feynman:1996kb},
delivered in 1962-63 and aimed to advanced graduate students and
postdocs. Finally, there is some unpublished material, included in
two sets of lectures, given in the 1960s at the Hughes Aircraft
Company, which only recently have been made available on the
web.\footnote{See Ref. \citenum{Feynmanbio} for a brief account of
Feynman's involvement in teaching at the Hughes Aircraft Company.}
In particular, the 1966-67 set of lectures \cite{FeynmanHughes1},
which were devoted to astronomy, astrophysics and cosmology,
contains an introductory treatment of general relativity, with an
emphasis on applications to the main subjects. While sharing many
similarities with the above quoted Caltech lectures
\cite{Feynman:1996kb}, the Hughes treatment offers to the
attentive reader several original points. In those same years
Feynman had succeeded in finding a new derivation of Maxwell's
equations \cite{DeLuca:2019ija,DiMauro:2020bpd}, and a
generalization of this approach to gravity is suggested (but not
pursued) in several places in the Hughes lectures on astrophysics,
as well as in the set of lectures given in the following year and
devoted to electromagnetism \cite{FeynmanHughes2}.

After outlining the main steps and sources which helped us to
reconstruct the full development of Feynman's work on
gravitation,\footnote{See Ref. \citenum{noi} for a comprehensive
account of this work.} in this contribution we focus on two key
issues: the formulation of quantum gravity as a quantum field
theory of a massless spin-$2$ field, the graviton, in whole
analogy with quantum electrodynamics, which is the content of
Section 2, and the unitarity and renormalization issues arising
beyond the tree level approximation, presented in Section 3. Our
concluding remarks close the paper.

\section{Gravity as a quantum field theory}

Feynman's strategy in approaching gravity was firstly outlined at
Chapel Hill conference in a series of critical comments (Ref.
\citenum{ChapelHill}, pp. 272-276), in which a non-geometric and
field theoretical line of attack is put forward. His starting
point was an hypothetical, counterfactual situation, in which
scientists would discover the principles of Lorentzian quantum
field theories before general relativity.\footnote{In fact, in
Ref. \citenum{ChapelHill}, p. 273, Feynman said: ``Instead of
trying to explain the rest of physics in terms of gravity I
propose to reverse the problem by changing history. Suppose
Einstein never existed [...]".} The main concern in such a
situation would then be to include a new force, the gravitational
one, in the framework of quantum field theory. This approach would
be later completed in the first part of the Caltech lectures
\cite{Feynman:1996kb}. Feynman's reasoning was the following: on
the basis of the general principles of quantum field theory and of
experimental results it is possible to conclude that gravity, as
any other force, has to be mediated by exchanges of a virtual
particle, which in this case is a massless neutral spin-$2$
quantum, the graviton. Thus, by constructing a Lorentz invariant
quantum field theory of the graviton\footnote{The linear theory
for a massless spin-$2$ field and its massive counterpart was
completely worked out by Markus Fierz and Wolfgang E. Pauli in
Ref. \citenum{Fierz:1939ix}, on the basis of a previous work by
Paul A. M. Dirac \cite{Dirac:1936tg}, while iterative arguments
similar to Feynman's ones were later put forward by Suraj N. Gupta
\cite{Gupta:1954zz} and Robert H. Kraichnan
\cite{Kraichnan:1955zz} in an attempt to generate infinite
nonlinear terms both in the Lagrangian and in the stress-energy
tensor. See for details Ref. \citenum{LectGravPreface}.} and by
imposing certain consistency requirements, full general relativity
should be recovered. Clearly, by following the same procedure for
the spin-$1$ case Maxwell's equations are obtained (in this case
it is much simpler, being the theory linear). Such an approach testifies his
ideas about fundamental interactions as manifestations of
underlying quantum theories \cite{DiMauro:2020bpd}, which were
expressed by him several times, for example in the Hughes
Lectures\cite{FeynmanHughes2}:
\begin{quote}
I shall call conservative forces, those forces which can be
deduced from quantum mechanics in the classical limit. As you
know, Q.M. is the underpinning of Nature (Ref.
\citenum{FeynmanHughes2}, p.35).
\end{quote}
Let us describe the steps in more detail. First of all, one has to
establish the spin of the mediating quantum. Both in the Caltech
\cite{Feynman:1996kb} and in the 1966-67 Hughes lectures
\cite{FeynmanHughes1}, the choice of a spin-$2$ mediator was
justified by the observation that energy, which is the source of
the gravitational force, grows with the velocity. The same
observation ruled out the possibility of a spin-$0$ field, because
the associated charge would decrease with the velocity. This
result can be traced back to an old argument by Einstein (never
published but recalled in Ref. \citenum{Einstein1933}, pp.
285-290), according to which the vertical acceleration of a body
would depend on its horizontal velocity, and in particular would
be zero for light, making light deflection impossible. Once the
spin of the graviton is established, one can easily construct the
linearized theory of the associated field, which is a massless
spin-$2$ field. Nonlinearity then comes into play because the
graviton has to couple with anything carrying energy-momentum,
then also with itself, and this coupling must be universal. The
resulting nonlinear theory is general relativity. General
covariance and the geometric interpretation of general relativity
are finally recovered as a byproduct of the gauge invariance of
the theory \footnote{See Ref. \citenum{Feynman:1996kb}, p. 113.}.
For Feynman, this was only half of the whole story. In fact, by
pushing calculations beyond tree level, quantum gravity effects
would be taken into account. This was Feynman's ultimate goal,
i.e. obtaining a quantum theory of gravity, which in this approach
amounts to the quantization of another field.

Let us describe how Feynman sketched the above procedure at the
Chapel Hill conference (Ref. \citenum{ChapelHill}, pp. 272-276).
In whole analogy with electrodynamics, he wrote down the following
action:
\begin{eqnarray}\label{secondOrAction}
\int \left(\frac{\partial A_{\mu}}{\partial x_{\nu}}-\frac{\partial A^{\nu}}{\partial x^{\mu}}\right)^2 {\rm d}^4x + \int A_{\mu}j^{\mu} {\rm d}^4x + \frac{m}{2}\int \dot{z}_{\mu}^2 {\rm d} s + \frac{1}{2}\int T_{\mu\nu}h^{\mu\nu} {\rm d}^4x \nonumber \\
+ \int (\textrm{second power of first derivatives of $h$} ),
\end{eqnarray}
where $h_{\mu\nu}$ is the new field associated with the graviton,
i.e. a symmetric second-order tensor field, which satisfies second
order linear equations of the kind:
\begin{eqnarray} \label{fieldEq}
{h^{\mu\nu,\sigma}}_{,\sigma} - 2
{{\overline{h}^{\mu}}_{\sigma}}^{,\nu\sigma}=
\overline{T}^{\mu\nu},
\end{eqnarray}
where the {\it bar} operation is defined on a generic second rank
tensor $X_{\mu\nu}$ as:
\begin{eqnarray} \label{barOperation}
\overline{X}_{\mu\nu}=\frac{1}{2} \left( X_{\mu\nu} + X_{\nu\mu}
\right) - \frac{1}{2} \eta_{\mu\nu} {X^{\sigma}}_{\sigma}.
\end{eqnarray}
The equations of motion for particles also follow from the above
action:
\begin{eqnarray} \label{particleEq}
g_{\mu\nu} \ddot{z}^{\nu} = -\left[\rho\sigma,\mu
\right]\dot{z}^{\rho}\dot{z}^{\sigma},
\end{eqnarray}
where $g_{\mu\nu}=\eta_{\mu\nu}+h_{\mu\nu}$, $\eta_{\mu\nu}$ is
the Minkowski metric, and $\left[\rho\sigma,\mu \right]$ are the
Christoffel symbols of the first kind.

However, when the field $h_{\mu\nu}$ is coupled to the matter
according to Eq.(\ref{particleEq}), the corresponding stress
energy tensor $T_{\mu\nu}$ does {\it not} obey to a conservation
law, leading to a consistency problem. This happens at variance
with electromagnetism, where Maxwell equations guarantee
conservation of the current $j^{\mu}$. Instead here $T_{\mu\nu}$
does not take into account the effect of gravity on itself, which
requires nonlinearity.\footnote{Feynman noticed that nonlinearity
was necessary in order to explain the precession of the perihelion
of Mercury (as discussed in Ref. \citenum{Feynman:1996kb}, p.
75).} Thus a suitable $T_{\mu\nu}$ has to be found in order to
satisfy the condition $\partial_{\nu}T^{\mu\nu}=0$. The solution
to this consistency issue\footnote{See for details Refs.
\citenum{ChapelHill}, p. 274 and \citenum{Feynman:1996kb}, pp.
78-79.} can be obtained by adding to the action a nonlinear third
order term in $h_{\mu\nu}$, which leads to the following equation
for $T_{\mu\nu}$:
\begin{eqnarray} \label{tensorEq}
g_{\mu\lambda} T^{\mu\nu}_{\;\;\;\;\,,\nu} =-
\left[\rho\nu,\lambda \right]T^{\rho\nu}.
\end{eqnarray}
One can then go on to higher order approximations, until the
procedure converges. But finding the general solution of Eq.
(\ref{tensorEq}) is a really difficult task in the absence of a
standard procedure. Feynman's idea was to look for an expression
for the $T_{\mu\nu}$, and hence for the action, that is invariant
under the following infinitesimal transformation of the whole
tensor field $g_{\mu\nu}$:
\begin{eqnarray} \label{infinitesimal1}
g_{\mu\nu}' = g_{\mu\nu} + g_{\mu\lambda}\frac{\partial
A^{\lambda}}{\partial x^{\nu}} + g_{\nu\lambda}\frac{\partial
A^{\lambda}}{\partial x^{\mu}} + A^{\lambda}\frac{\partial
g_{\mu\nu}}{\partial x^{\lambda}}.
\end{eqnarray}
where the $4-$vector $A^{\lambda}$ is the generator. This is a
geometric transformation in a Riemannian manifold with metric,
hence one can say that geometry gives the metric $g_{\mu\nu}$. As
such, in Feynman's approach geometry comes into play at the end
and not at the beginning. By working out calculations, the full
nonlinear Einstein gravitational field equations are obtained.

Lectures 3-6 of Feynman's graduate course on gravitation (Ref.
\citenum{Feynman:1996kb}) contain all the details of the above
procedure. Interestingly, a proof is also given of the ability of
this field theory based formalism to reproduce key physical
effects of curved spacetime geometry. For instance, in Lecture 5
(Ref. \citenum{Feynman:1996kb}, pp. 66-69) it is shown that, in
the action of a scalar field, the time dilation $t\rightarrow
t'=t\sqrt{1+\epsilon}$ exactly reproduces the effect of a constant
weak gravitational field described by the tensor
$g_{44}=1+\epsilon$, $g_{ii}=-1$, $i=1,2,3$. Incidentally such an
effect plays a pivotal role in producing the right result for the
precession of Mercury's perihelion. Before moving to applications,
Feynman devoted some lectures to the discussion of the usual
geometric approach to gravity and of its link with the field
theory based approach (Ref. \citenum{Feynman:1996kb}, Lectures
7-10):
\begin{quote}
Let us try to discuss what it is that we are learning in finding
out that these various approaches give the same results (Ref.
\citenum{Feynman:1996kb}, p. 112).
\end{quote}
Despite advocating one approach over another, Feynman was in fact
intrigued by the double nature of gravity, which has both a
geometrical interpretation and a field interpretation, and in
Section 8.4 of Ref. \citenum{Feynman:1996kb} he recognized how an
explanation may be provided by gauge invariance. Indeed a viable
procedure may be established in order to obtain the invariance of
the equations of physics under space dependent variable
displacements. This amounts to looking for a more general
Lagrangian, to be obtained by adding to the old one new terms,
involving a gravity field. The net result is a new picture of
gravity, as the field corresponding to a gauge invariance with
respect to displacement transformations.

Summing up, Feynman succeeded in obtaining the full nonlinear
Einstein equations by means of a consistency argument applied to a
Lorentz invariant quantum field theory of the graviton. According
to John Preskill and Kip S. Thorne \cite{LectGravPreface}, it is
likely that he was completely unaware of the earlier work by Gupta
\cite{Gupta:1954zz} and Kraichnan \cite{Kraichnan:1955zz}, which
as mentioned had been developed along a similar line of attack,
but was still incomplete. So he would have developed his approach
independently, besides getting more complete results. A
Lorentz-invariant field theoretical approach to gravity would have
been later pursued also by Steven Weinberg
\cite{Weinberg:1964kqu,Weinberg:1965rz}, albeit quite different
from Feynman's one,\footnote{Weinberg's approach relied on the
analyticity properties of graviton-graviton scattering amplitudes,
and it was quite more general, since Weinberg actually proved that
a quantum theory of a massless spin-2 field can only be consistent
if this field universally couples to the energy-momentum of
matter, hence the equivalence principle has to be obeyed.} as well
as by Stanley Deser
\cite{Deser:1969wk,Boulware:1974sr,Deser87,Deser:2009fq}. Indeed
Deser's approach, while being similar to Feynman's one, was more
elegant and general and led to completion the whole program
started with Kraichnan and Gupta. Finally, a rigorous and general
analysis of the relation between spin-$2$ theories and general
covariance was carried out by Robert M. Wald \cite{Wald:1986bj}.

\section{Fighting with loops: the renormalization of gravity}
After obtaining the Einstein-Hilbert action and the full nonlinear
Einstein gravitational field equations, Feynman's efforts were
mainly directed towards discussing quantum field theory issues
beyond the tree level approximation, i.e. loop diagrams, unitarity
and renormalization.

To the best of our knowledge, these issues were publicly addressed
for the first time by Feynman in 1961 at the already mentioned La
Jolla conference \cite{LaJolla}, where nonlinearity was recognized
as the very source of difficulty within both gravitation and
Yang-Mills theories. Indeed the sources of the gravitational field
are energy and momentum, and the gravitational field carries
energy and momentum itself. In the same way, the source of a
Yang-Mills field is the isotopic spin current, and the Yang-Mills
field carries isotopic spin itself. This means that both the
gravitational field and the Yang-Mills field are self-coupled,
resulting in nonlinear field theories. A difficulty that
nonlinearity brings about is the fact that loop diagrams seem to
clash with unitarity.

As recalled in the introduction, Feynman's results on quantum
gravity can be found in a report of the talk given at the 1962
Warsaw conference, later published in Acta Physica Polonica
\cite{Feynman:1963ax},\footnote{As pointed out by Trautman in some
recently published memories (\cite{Trautman}, p. 406), the text of
Feynman's plenary lecture became available too late to be included
in the proceedings, therefore it was published only in 1963 as a
regular paper.} with many details discussed much later in the two
1972 Wheeler Festschrift papers \cite{WheelerFest1,WheelerFest2}.

Also in this case Feynman followed his original strategy, leaving
aside quantization of space-time geometry, constructing a quantum
field theory for the graviton, and working out results at
different perturbative orders. Since the goal was the quantum
theory, the Einstein equations and the corresponding Lagrangian
were assumed as a starting point, rather than derived from
scratch. The theory was coupled to a scalar field, and
perturbative calculations up to the next to leading order were
pursued. This implied the inclusion of loop diagrams, which make
quantum corrections to enter the game. In Feynman's own words:
\begin{quote}
I started with the Lagrangian of Einstein for the interacting
field of gravity and I had to make some definition for the matter
since I'm dealing with real bodies and make up my mind what the
matter was made of; and then later I would check whether the
results that I have depend on the specific choice or they are more
powerful (Ref. \citenum{Feynman:1963ax}, p. 698).
\end{quote}
The metric was split in the following way:
\begin{eqnarray}\label{metric}
g_{\mu\nu} = \delta_{\mu\nu} + \kappa h_{\mu\nu};
\end{eqnarray}
where the Minkowski metric is here denoted by $\delta_{\mu\nu}$
and $\kappa$ is a dimensionful coupling constant. Substituting
(\ref{metric}) and expanding, the Lagrangian for gravity coupled
with a scalar field can be cast in the form:
\begin{eqnarray} \label{lagrangian}
L &=& \int  \left(h_{\mu\nu,\sigma} \overline{h}_{\mu\nu,\sigma} - 2 \overline{h}_{\mu\sigma,\sigma}\overline{h}_{\mu\sigma,\sigma}\right) {\rm d} \tau + \frac{1}{2} \int \left( \phi_{,\mu}^2 - m^2 \phi^2 \right) {\rm d} \tau \\
& & + \kappa \int
\left(\overline{h}_{\mu\nu}\phi_{,\mu}\phi_{,\nu} - m^2
\frac{1}{2}{h}_{\sigma\sigma}\phi^2 \right) {\rm d} \tau + \kappa
\int \left( hhh \right){\rm d} \tau + \kappa^2 \int \left(
hh\phi\phi \right) {\rm d} \tau + ...\nonumber
\end{eqnarray}
where the bar operation has been defined in (\ref{barOperation})
and a schematic notation has been adopted for the highly complex
higher order terms. The first two terms are simply the free
Lagrangians of the gravitational field and of matter,
respectively. Before considering radiative corrections the
classical solution of the problem was worked out, which involved
the variation of Eq. (\ref{lagrangian}) with respect to $h$ and,
then, to $\phi$, giving rise to the following equations of motion
with a source term:
\begin{eqnarray} \label{source1}
h_{\mu\nu,\sigma\sigma} - \overline{h}_{\sigma\nu,\sigma\mu} -
\overline{h}_{\sigma\mu,\sigma\nu} &=& \overline{S}_{\mu\nu}
\left(h, \phi \right),\\ \label{source2} \phi_{,\sigma\sigma} -
m^2 \phi &=& \chi \left(\phi, h \right).
\end{eqnarray}
A close inspection revealed that Eq. (\ref{source1}) was singular,
so that Feynman was forced to resort to the invariance of the
Lagrangian under the transformation:
\begin{eqnarray} \label{invariance}
h_{\mu\nu}^{'} = h_{\mu\nu} + 2 \xi_{\mu,\nu} + 2 h_{\mu\sigma}\xi_{\sigma,\nu} + \xi_{\sigma} h_{\mu\nu,\sigma},
\end{eqnarray}
$\xi_{\mu}$ being arbitrary. This meant that the source
$S_{\mu\nu}$ had to be divergenceless in order to make Eq.
(\ref{source1}) consistent. Finally, by making the gauge choice
$\overline{h}_{\mu\sigma,\sigma}=0$, the law of the gravitational
interaction of two systems by means of the exchange of a virtual
graviton was obtained. Feynman then went on computing other
processes, such as an interaction vertex coupling two particles
and a graviton and the gravitational analog of gravitational
Compton effect (i.e. with a graviton replacing the photon).

After these preliminary calculations, Feynman went to the next to
leading order approximation, thus encountering diagrams with
closed loops:
\begin{quote}
However the next step is to take situations in which we have what
we call closed loops, or rings, or circuits, in which not all
momenta of the problem are defined (Ref. \citenum{Feynman:1963ax},
pp. 703-704).
\end{quote}
He realized that working out closed loop diagrams required the
solution of a number of conceptual issues, and he succeeded in
showing that any diagram with closed loops can be expressed in
terms of sums of on shell tree diagrams, which is the content of
his celebrated tree theorem (which was treated in detail in Ref.
\citenum{WheelerFest1}). Further details on the statement of the
tree theorem and, in particular, on the nature of the proof for
the one-loop case were given by Feynman in the discussion section
(Ref. \citenum{Feynman:1963ax}, pp. 714-717), while answering some
related questions by DeWitt. But the main problem to face in
carrying out one-loop calculations was the lack of unitarity, due
to the presence of contributions arising from the unphysical
longitudinal polarization states of the graviton, which did not
cancel as they should. Following a suggestion by
Gell-Mann,\footnote{``I suggested that he try the analogous
problem in Yang-Mills theory, a much simpler nonlinear gauge
theory than Einsteinian gravitation.'' (Ref. \citenum{Gell-Mann},
p. 53).} Feynman considered the simpler Yang-Mills case (his
results in this case were summarized in Ref.
\citenum{WheelerFest2}) and found the same pathological behavior:
\begin{quote}
But this disease which I discovered here is a disease which exist
in other theories. So at least there is one good thing: gravity
isn't alone in this difficulty. This observation that Yang-Mills
was also in trouble was of very great advantage to me. [...] the
Yang-Mills theory is enormously easier to compute with than the
gravity theory, and therefore I continued most of my
investigations on the Yang-Mills theory, with the idea, if I ever
cure that one, I'll turn around and cure the other (Ref.
\citenum{Feynman:1963ax}, p. 707).
\end{quote}
The solution to this issue was obtained by expressing each loop
diagram as a sum of trees and then computing the trees. This
worked even if the process of opening a loop by cutting a graviton
line implies the replacement of a virtual graviton with a real
transverse one. Finally, in order to guarantee gauge invariance
the sum of the whole set of tree diagrams corresponding to a given
process has to be taken.

The same results, according to Feynman, could be obtained by
direct integration of the closed loop. In the last case a
mass-like term has to be added to the Lagrangian to avoid
singularity but at the price of breaking gauge invariance. At the
same time a contribution has to be subtracted, which is obtained
by making a ghost particle (with spin-$1$ and Fermi statistics) to
go around the loop and artificially coupled to the external field.
In this way both unitarity and gauge invariance would be restored.
This procedure was worked out also for Yang-Mills theory, but in
that case the ghost particle has spin-$0$ \cite{WheelerFest2}.

Once successfully solved the one-loop case, Feynman's efforts
pointed toward a further generalization of the above procedure to
two or more loops:
\begin{quote}
Now, the next question is, what happens when there are two or more
loops? Since I only got this completely straightened out a week
before I came here, I haven't had time to investigate the case of
2 or more loops to my own satisfaction. The preliminary
investigations that I have made do not indicate that it's going to
be possible so easily gather the things into the right barrels.
It's surprising, I can't understand it; when you gather the trees
into processes, there seems to be some loose trees, extra trees
(Ref. \citenum{Feynman:1963ax}, p. 710).
\end{quote}
But, in Feynman's words, preliminary attempts seem to suggest that
novel difficulties enter the game when dealing with two or more
loops, as also mentioned in the last of his published Lectures on
Gravitation (Ref. \citenum{Feynman:1996kb}, Lecture 16, pp.
211-212). Here, once again, he recognized in the lack of unitarity
of some sums of diagrams the main source of the observed
pathological behavior and pointed out that a similar feature was
shared also by Yang-Mills theory. Finally, while hinting at the
problem of finding ghost rules for high order diagrams, he argued
in favor of the non-renormalizability of gravity as a consequence
of these difficulties:
\begin{quote}
I do not know whether it will be possible to develop a cure for
treating the multi-ring diagrams. I suspect not -- in other words,
I suspect that the theory is not renormalizable. Whether it is a
truly significant objection to a theory, to say that it is not
renormalizable, I don't know (Ref. \citenum{Feynman:1996kb},
Lecture 16, pp. 211-212).
\end{quote}
It is not clear whether Feynman was suggesting a link between
non-unitarity and non-renormalizability issues. But in any case
Feynman's results played a prominent role in the development of
gauge theory and quantum gravity. Feynman's rules for ghosts were
later generalized to all orders by DeWitt \cite{DeWitt:1964yg,
DeWitt:1967ub, DeWitt:1967uc}, while Ludvig D. Faddeev and Viktor
N. Popov \cite{FaddeevPopov} derived them in a much simpler way,
by means of functional integral quantization, setting the standard
for all subsequent work in the field. In particular, DeWitt proved
that Yang-Mills theory and quantum gravity are in fact unitary at
two \cite{DeWitt:1964yg} and arbitrarily many loops
\cite{DeWitt:1967ub, DeWitt:1967uc}. However, while Yang-Mills
theory was later shown to be renormalizable (cf. Refs.
\citenum{tHooft:1971akt}-\citenum{tHooft:1972qbu}), gravity
presented divergences which could not be renormalized (cf. Refs.
\citenum{tHooft:1974toh}-\citenum{vandeVen:1991gw}), confirming
Feynman's suspect. It should be mentioned that, in subsequent
years, modified theories of gravity, characterized by an action
quadratic in the curvature, have been put forward. Unlike ordinary
general relativity, these theories are renormalizable but not
unitary \cite{Stelle:1976gc}.

It is worth mentioning that, unlike most of his contemporaries,
Feynman did not think about non-renormalizability as a signature
of inconsistency of a theory, as also recalled by
Gell-Mann,\footnote{``He was always very suspicious of
unrenormalizability as a criterion for rejecting theories'' (Ref.
\citenum{Gell-Mann}, p. 53).} and claimed by Feynman himself in
one of his last interviews, given in January 1988:
\begin{quote}
The fact that the theory has infinities never bothered me quite so
much as it bother others, because I always thought that it just
meant that we've gone too far: that when we go to very short
distances the world is very different; geometry, or whatever it
is, is different, and it's all very subtle (\cite{Mehra:1994dz}, p. 507).
\end{quote}
In fact, within the modern view on quantum field theory, which was
developed in 1970s, non-renormalizability is considered only a
signature of the fact that the theory loses its validity at
energies higher that a certain scale. Nevertheless, one has an
\emph{effective field theory}, which works and can be useful to
make predictions under that scale (interesting historical
discussions can be found in Refs. \citenum{Weinberg:2021exr} and
\citenum{Weinberg:2016kyd}). This is true also for gravity
\cite{Burgess:2003jk}. But, as remembered by John P. Preskill in a
recent talk (Ref. \citenum{PreskillTalk}, slide 37), although he
anticipated this view, apparently Feynman was not really at ease
with it:
\begin{quote}
I spoke to Feynman a number of times about renormalization theory
during the mid-80s (I arrived at Caltech in 1981 and he died in
1988). I was surprised on a few occasions how the effective field
theory viewpoint did not come naturally to him''. [...] Feynman
briefly discusses in his lectures on gravitation (1962) why there
are no higher derivative terms in the Einstein action, saying this
is the ``simplest'' theory, not mentioning that higher derivative
terms would be suppressed by more powers of the Planck length.
\end{quote}
As a further remark, let us notice that in the last years
Feynman's tree theorem has spurred a renewed interest in
researchers working in the context of advanced perturbative
calculations and generalized unitarity (cf. Refs.
\citenum{Britto:2010xq} - \citenum{Maniatis:2019pig}).

\section{Concluding remarks}
In this paper we focused on Feynman's contributions to the
research in quantum gravity, starting from his interventions at
Chapel Hill conference in 1957 and ending with the Wheeler
festschrift papers. His approach was field theoretical rather than
geometric, reflecting his strong belief in the unity of Nature,
which is quantum at the deepest level. Quantization of gravity,
according to him, simply had to be considered as the quantization
of another field, the spin-$2$ graviton field. In this way full
general relativity would be recovered at leading order, while the
inclusion of loop diagrams brought into the picture a bunch of new
difficulties. In this respect, Feynman's struggle against loops,
while succeeding at one-loop order, failed with two- and
higher-loop diagrams. Nevertheless, his results triggered further
efforts and some tools he developed, such as the tree theorem,
have recently become of widespread use among people working on
scattering amplitudes.




\end{document}